\documentclass[10pt,twocolumn,aps,pra,amsmath,amssymb,showpacs]{revtex4-1}
\usepackage{bm}
\usepackage{mathrsfs}
\usepackage{graphicx}
\usepackage[usenames,dvipsnames,svgnames,table]{xcolor}
\usepackage{framed}
\usepackage{amsthm}
\usepackage[utf8]{inputenc}

\colorlet{shadecolor}{blue!15}
\usepackage[unicode=true,
            pdfusetitle, 
            bookmarks=true,
            bookmarksnumbered=false,
            bookmarksopen=false,
            breaklinks=true,
            pdfborder={0 0 0},
            backref=false,
            colorlinks=true]{hyperref}
\hypersetup{linkcolor=NavyBlue,urlcolor=NavyBlue,citecolor=NavyBlue}

\usepackage{amsthm}
\newcommand{\propnumber}{} 
\newtheorem*{prop}{Proposition \propnumber}
\newenvironment{propc}[1]
  {\renewcommand{\propnumber}{#1}%
  \begin{prop}}
  {\end{prop}}
  \newtheorem{propo}{Proposition} 
 \newtheorem{rem}{Remark}

 \usepackage{tikz-cd}

\newcommand{\cH}{\mathcal{H}}

\begin{document}

\title {Necessary and sufficient condition for the reduced dynamics of an open quantum
system interacting with an environment to be linear}

\author {Iman Sargolzahi}
\email{sargolzahi@neyshabur.ac.ir; sargolzahi@gmail.com}
\affiliation {Department of Physics, University of Neyshabur, Neyshabur, Iran}

\affiliation{Research Department of Astronomy and Cosmology, University of Neyshabur, Neyshabur, Iran}

\begin{abstract}
The dynamics of a closed quantum system, under a unitary time evolution $U$, is, obviously, linear. But, the reduced dynamics of an open quantum system $S$, interacting with an environment $E$, is not linear, in general. 
  Dominy \textit{et al}.   [J. M. Dominy, A. Shabani, and D. A. Lidar,    \href{http://link.springer.com/article/10.1007/s11128-015-1148-0}
    { Quantum Inf. Process. {\bf 15}, 465 (2016)}] 
considered the case that  the set $\mathcal{S}=\lbrace\rho_{SE}\rbrace$, of possible initial states of the system-environment, is convex and, also, possesses another property, which they called $U$-\textit{consistency}.  They have shown that, under such circumstances, the reduced dynamics of the system $S$ is linear.
Whether the   Dominy-Shabani-Lidar  framework is the most general one is the subject of this paper. 
We assume that the reduced dynamics is linear and show that this leads us to their 
 framework. In other words, the reduced dynamics of the system is linear if and only if it can be formulated within  the   Dominy-Shabani-Lidar  framework.
\end{abstract}

\maketitle

\section{Introduction}

The time evolution of a closed quantum system is given by
\begin{equation}
\label{eq:1}
\begin{aligned}
\rho^{\prime}= \mathrm{Ad}_{U}(\rho)\equiv  U \rho U^{\dagger}, 
\end{aligned}
\end{equation} 
where $\rho$ and  $\rho^{\prime}$ are the initial and the final states (density operators) of the system, respectively, and $U$ is a unitary operator \cite{1}. 
When the system $S$ is not closed and interacts with its environment $E$, we can consider the whole system-environment as a closed quantum system, which evolves as Eq. \eqref{eq:1}, and so, the reduced dynamics of the system is given by
\begin{equation}
\label{eq:2}
\begin{aligned}
\rho_{S}^{\prime}=\mathrm{Tr}_{E} \circ \mathrm{Ad}_{U}(\rho_{SE})= \mathrm{Tr}_{E}\left( U \rho_{SE}U^{\dagger}\right), 
\end{aligned}
\end{equation} 
where $\rho_{S}^{\prime}$ is the final state of the system, $\rho_{SE}$ is the initial state of the system-environment,  and the unitary operator $U$ acts on the whole Hilbert space of the system-environment \cite{1}.

 The initial state of the system is $\rho_{S}=\mathrm{Tr}_{E} (\rho_{SE})$. An important question, in the theory of open quantum systems \cite{2}, is  whether there exists a map $\Phi_S$ such that
\begin{equation}
\label{eq:3}
\rho_{S}^{\prime}=\Phi_S(\rho_{S}),
\end{equation} 
i.e., whether the final state $\rho_{S}^{\prime}$ can be written as a function of the initial state $\rho_{S}$. In general, it is not the case \cite{ 3, 4, 2}.  Even if there exists such a map,  $\Phi_S$ is not linear, in general \cite{5, 6}. 

However, if there exists a linear map  $\Phi_S$, then it can be shown that this  \textit{dynamical map} $\Phi_S$ is, in addition, Hermitian, \cite{7, 8}, i.e., maps each Hermitian operator to a Hermitian operator.
For each linear trace-preserving Hermitian map  $\Phi_S$,  there exists an operator sum representation  as
\begin{equation}
\label{eq:4}
\begin{aligned}
\rho_{S}^{\prime}=\Phi_S(\rho_{S})=\sum_{i}e_{i}\,\tilde{E_{i}}\,\rho_{S}\,\tilde{E_{i}}^{\dagger},\ \\  \sum_{i}e_{i}\,\tilde{E_{i}}^{\dagger}\tilde{E_{i}}=I_{S}, \qquad\qquad
\end{aligned}
\end{equation}
where $\tilde{E_{i}}$ are linear operators and $I_S$ is the identity operator, on the Hilbert space of the system $\cH_S$, and $e_{i}$ are real coefficients \cite{7, 8}.

For the special case that all of the coefficients $e_{i}$ in Eq. \eqref{eq:4} are positive,  we can define $E_{i}=\sqrt{e_{i}}\,\tilde{E_{i}}$, and so Eq. \eqref{eq:4} can be rewritten as
\begin{equation}
\label{eq:5}
\begin{aligned}
\rho_{S}^{\prime}=\Phi_S(\rho_{S})=\sum_{i} E_{i}\,\rho_{S}\,E_{i}^{\dagger},\ \ \ \ \ 
  \sum_{i} E_{i}^{\dagger}E_{i}=I_{S}. 
\end{aligned}
\end{equation}
A map  $\Phi_S$, which can be written as  Eq. \eqref{eq:5}, is called a completely positive map \cite{1, 2}.

 An important question remains: When can the reduced dynamics be given by a linear map $\Phi_S$?
In Ref. \cite{4}, Dominy \textit{et al}. considered the case that the set $\mathcal{S}=\lbrace\rho_{SE}\rbrace$, of possible initial states of the system-environment, is convex.  They have shown that, if, in addition,  $\mathcal{S}$ possesses a necessary condition, which they called $U$-\textit{consistency}, the reduced dynamics is linear. In the next section, we will review their framework.

Investigating whether their framework is the most general one is the subject of this paper. So, we assume that the reduced dynamics is linear and show that this assumption leads us to their framework. Therefore, the reduced dynamics of the system is given by a linear map $\Phi_S$  if and only if it can be formulated within  their  framework. This result, as our main result, is given in Sec. \ref{sec: C}. 

In Sec. \ref{sec: D}, we illustrate our result, studying an example, given in Ref. \cite{8}. We  discuss whether the nonlinearity of the reduced dynamics results in superluminal signaling, or not, in Sec. \ref{sec: E}. 
 Finally, we end our paper in Sec.  \ref{sec: F}, with a summary of our results.

\section{Dominy-Shabani-Lidar  framework for the reduced dynamics}  \label{sec: B}

A general framework for the linear Hermitian trace-preserving reduced dynamics, when both the system $S$ and the environment $E$ are finite dimensional, has been introduced in \cite{4}.
This framework can be easily generalized to the case that the system is finite dimensional, with the dimension $d_S$, but the dimension of the environment is arbitrary, i.e., $E$ can be infinite  dimensional \cite{9}. In this section, we review this generalized version of the Dominy-Shabani-Lidar framework, in such a way that helps us achieve our main result, in the next section.

Consider the set $\mathcal{S}=\lbrace\rho_{SE}\rbrace$ of possible initial states of the system-environment. So, the set of possible initial states of the system is given by $\mathcal{S}_S=\mathrm{Tr}_E\mathcal{S}$.
 Since  the system $S$ is finite dimensional, a finite number $m$ of the members of $\mathcal{S}_S$, where the integer $m$ is  $0< m\leq {(d_S)}^2$, are linearly independent. Let us denote this linearly independent set as
  $\mathcal{S}^\prime_S  = \lbrace\rho_{S}^{(1)}, \rho_{S}^{(2)}
  , \ldots\ , \rho_{S}^{(m)}\rbrace$. Therefore, any $\rho_{S}\in \mathcal{S}_S$ can be expanded as
\begin{equation}
\label{eq:6}
\begin{aligned}
 \rho_{S}=\sum_{i=1}^m a_i \rho_{S}^{(i)},
\end{aligned}
\end{equation}  
   where $a_i$ are real coefficients.

Linear independence of   $\rho_{S}^{(i)}\in \mathcal{S}^\prime_S$ results in linear independence of  $\rho_{SE}^{(i)}$, where $\rho_{S}^{(i)}=\mathrm{Tr}_E(\rho_{SE}^{(i)})$. We denote this linearly independent set as   $\mathcal{S}^\prime  = \lbrace\rho_{SE}^{(1)}, \rho_{SE}^{(2)}
  , \ldots\ , \rho_{SE}^{(m)}\rbrace$. So, each $\rho_{SE}\in \mathcal{S}$ can be written as 
 \begin{equation}
\label{eq:7}
\begin{aligned}
 \rho_{SE}=\sum_{i=1}^m a_i \rho_{SE}^{(i)}+Y,
\end{aligned}
\end{equation}  
   where $a_i$ are the same as those  in Eq. \eqref{eq:6},  and $Y$ is a Hermitian operator (on $\cH_S\otimes\cH_E$, where $\cH_E$ is the Hilbert space of the environment), such that $\mathrm{Tr}_E (Y)=0$.
 Equation \eqref{eq:7} means that if we cannot expand $ \rho_{SE}$ by 
  $\rho_{SE}^{(i)}\in \mathcal{S}^\prime$, then, since  $\rho_{S}=\mathrm{Tr}_E (\rho_{SE})$ is given by Eq. \eqref{eq:6}, the difference between $ \rho_{SE}$ and $\sum_{i=1}^m a_i \rho_{SE}^{(i)}$ is a  $Y$, such that $\mathrm{Tr}_E (Y)=0$.

Now, if there exists another  $\tau_{SE}\in \mathcal{S}$, such that $\mathrm{Tr}_E (\tau_{SE})=\mathrm{Tr}_E (\rho_{SE})=\rho_{S}$, then, from Eqs. \eqref{eq:6} and \eqref{eq:7}, we have
 \begin{equation}
\label{eq:8}
\begin{aligned}
 \tau_{SE}=\sum_{i=1}^m a_i \rho_{SE}^{(i)}+\tilde{Y},
\end{aligned}
\end{equation}  
where $\mathrm{Tr}_E (\tilde{Y})=0$.
The  obvious requirement,  for the existence of  a map  $\Phi_S$  such that, for each $\rho_{S}=\mathrm{Tr}_E (\rho_{SE}) \in \mathcal{S}_S$, the final state $\rho_{S}^{\prime}=\mathrm{Tr}_{E} \circ \mathrm{Ad}_{U}(\rho_{SE})$  is given by $\Phi_S(\rho_{S})$, is the $U$-\textit{consistency} of the set $\mathcal{S}$ \cite{4}:  if for two  states in $\mathcal{S}$, e.g.,  $\rho_{SE}$ in Eq. \eqref{eq:7} and $\tau_{SE}$ in Eq. \eqref{eq:8}, we have $\mathrm{Tr}_E (\tau_{SE})=\mathrm{Tr}_E (\rho_{SE})=\rho_{S}$, then we must, also,  have 
 \begin{equation}
\label{eq:8a}
\begin{aligned}
\mathrm{Tr}_{E} \circ \mathrm{Ad}_{U}(\rho_{SE})=\mathrm{Tr}_{E} \circ \mathrm{Ad}_{U}(\tau_{SE}).
\end{aligned}
\end{equation}  
In other words, for both $\rho_{SE}$ and $\tau_{SE}$, the initial state of the system is the same (given by $\rho_{S}$), and so the final state of the system must be the same too, if we require that it is given by $\Phi_S(\rho_{S})$. This property is necessary and sufficient, for the existence of a map $\Phi_S$,  as Eq. \eqref{eq:3}.
 Using Eqs. \eqref{eq:7} and \eqref{eq:8},  the $U$-consistency property of the set  $\mathcal{S}$, in Eq. \eqref{eq:8a}, can be rewritten as
 \begin{equation}
\label{eq:9}
\begin{aligned}
\mathrm{Tr}_{E} \circ \mathrm{Ad}_{U}(Y- \tilde{Y}) =0.
\end{aligned}
\end{equation}

Next, let us define the subspaces $\mathcal{V}$ and $\mathcal{V}_S$ as \cite{4}
 \begin{equation}
\label{eq:10}
\begin{aligned}
\mathcal{V}= \mathrm{Span}_{\mathbb{C}} \  \mathcal{S},  
\end{aligned}
\end{equation}  
and 
\begin{equation}
\label{eq:11}
\begin{aligned}
\mathcal{V}_S=\mathrm{Tr}_{E} \mathcal{V}=\mathrm{Span}_{\mathbb{C}} \  \mathcal{S}_S=\mathrm{Span}_{\mathbb{C}} \  \mathcal{S}_S^\prime .
\end{aligned}
\end{equation}  
Therefore, each $X \in \mathcal{V}$ can be written as $X=\sum_{l} c_l \, \tau_{SE}^{(l)}$, where 
$\tau_{SE}^{(l)} \in  \mathcal{S}$, and $c_l$ are complex coefficients.
Using Eq. \eqref{eq:7}, we can expand each $\tau_{SE}^{(l)}$ as  $\tau_{SE}^{(l)}=\sum_{i} a_{li} \rho_{SE}^{(i)}+Y^{(l)}$. So,
\begin{equation}
\label{eq:12}
\begin{aligned}
 X=\sum_{i=1}^m  \left( \sum_{l}  a_{li} c_l \right) \rho_{SE}^{(i)}+\sum_{l} c_l \,  Y^{(l)} \\
 =\sum_{i=1}^m d_i \rho_{SE}^{(i)}+\hat{Y}, \qquad\qquad\qquad\quad
\end{aligned}
\end{equation}  
where $d_i=\sum_{l}  a_{li} c_l $ are complex coefficients, and the linear operator $\hat{Y}=\sum_{l} c_l \,  Y^{(l)}$ is such that $\mathrm{Tr}_E (\hat{Y})=0$. Consequently, for each $x\in \mathcal{V}_S$, we have
\begin{equation}
\label{eq:13}
\begin{aligned}
x =\mathrm{Tr}_E (X) =\sum_{i=1}^m d_i \rho_{S}^{(i)},
\end{aligned}
\end{equation}  
where the  coefficients $d_i$ are the same as those in Eq.  \eqref{eq:12}.

We have seen that the  $U$-consistency condition, for the set $\mathcal{S}$, results in the  existence of a map $\Phi_S$, such that Eq.  \eqref{eq:3} holds, for each $\rho_S \in \mathcal{S}_S$.
In the following, we will see that the  $U$-consistency property, for the subspace $\mathcal{V}$, results in the linearity of the map $\Phi_S$.

Consider the case that the subspace $\mathcal{V}$ is   $U$-consistent, 
 i.e., if, for $W,X\in  \mathcal{V}$, we have  $\mathrm{Tr}_E (W)=\mathrm{Tr}_E (X)=x$, then $\mathrm{Tr}_{E} \circ \mathrm{Ad}_{U}(W)=\mathrm{Tr}_{E} \circ \mathrm{Ad}_{U}(X)$.  
Note that, in Eq.
\eqref{eq:7}, both $\rho_{SE}$ and $\sum_{i=1}^m a_i \rho_{SE}^{(i)}$ are members of $\mathcal{V}$. So, the  $U$-consistency of $\mathcal{V}$ results in
\begin{equation}
\label{eq:14}
\begin{aligned}
\mathrm{Tr}_{E} \circ \mathrm{Ad}_{U}(Y)=0. 
\end{aligned}
\end{equation} 
The reverse is also true: if, for any $\rho_{SE}\in \mathcal{S}$, Eq. \eqref{eq:14} is satisfied, then, for $\hat{Y}$ in Eq.  \eqref{eq:12}, $\mathrm{Tr}_{E} \circ \mathrm{Ad}_{U}(\hat{Y})=0$, which means that  $\mathcal{V}$  is $U$-consistent. Therefore, $\mathcal{V}$  is $U$-consistent if and only if Eq. \eqref{eq:14} is satisfied, for any $\rho_{SE}\in \mathcal{S}$.
(Compare with the  $U$-consistency condition, for the set $\mathcal{S}$, in Eq. \eqref{eq:9}.)

Now, we define the linear trace-preserving \textit{assignment map} $ \Lambda_S $, as follows:  
for any $x\in \mathcal{V}_S$, in Eq. \eqref{eq:13}, we define
\begin{equation}
\label{eq:15}
\begin{aligned}
\Lambda_S(x)=\sum_{i=1}^m d_i \Lambda_S(\rho_{S}^{(i)})=\sum_{i=1}^m d_i \rho_{SE}^{(i)}.
\end{aligned}
\end{equation} 
The assignment map $ \Lambda_S $ maps $\mathcal{V}_S$ to (a subspace of)  $\mathcal{V}$, and is  Hermitian, by construction. (When $x$ is a Hermitian operator, all $d_i$ are real, and, obviously, $ \Lambda_S $ maps such a Hermitian $x$ to a Hermitian operator.)

Finally, using Eqs. \eqref{eq:2}, \eqref{eq:6}, \eqref{eq:7}, \eqref{eq:14} and \eqref{eq:15}, for each $\rho_{SE}\in \mathcal{S}$ (in fact, for each $\rho_{SE}\in \mathcal{V}$), we have 
\begin{equation}
\label{eq:16}
\begin{aligned}
\rho_{S}^{\prime}=\mathrm{Tr}_{E} \circ \mathrm{Ad}_{U}(\rho_{SE}) \qquad\qquad\qquad\qquad\qquad   \\
\qquad\quad = \sum_{i=1}^m a_i \mathrm{Tr}_{E} \circ \mathrm{Ad}_{U}(\rho_{SE}^{(i)})+\mathrm{Tr}_{E} \circ \mathrm{Ad}_{U}(Y) \\
=\mathrm{Tr}_{E} \circ \mathrm{Ad}_{U}  \circ \Lambda_S(\rho_{S})\equiv 
\Phi_S(\rho_{S}). \qquad\quad
\end{aligned}
\end{equation} 
The map $\Phi_S$ is  Hermitian, since $\mathrm{Tr}_{E}$ and $\mathrm{Ad}_{U}$ are completely positive \cite{1}, and 
 $ \Lambda_S $ is Hermitian. So,  $\Phi_S$ has an operator sum representation, as Eq. \eqref{eq:4}. 
 If $\Lambda_S$ is, in addition,  completely positive, then  $\Phi_S$ is so and has an  operator sum representation, as Eq. \eqref{eq:5}.
Whether there exists a completely positive $\Lambda_S$, or not, may be determined using the \textit{reference state} \cite{10, 11}. Nevertheless, it is also possible that $\Lambda_S$ is non-positive, but  $\Phi_S$ is completely positive \cite{4, 11}.

In summary, we have seen that if the subspace $\mathcal{V}$, in Eq. \eqref{eq:10},  is  $U$-consistent, then the reduced dynamics of the system is given by the 
linear (Hermitian trace-preserving) map  $\Phi_S$, in Eq. \eqref{eq:16}.
But, in the previous section, we have stated that if $\mathcal{S}$  is convex \cite{9-1} and $U$-consistent, then the reduced dynamics is linear. In fact, 
it can be shown that when $\mathcal{S}$  is convex and $U$-consistent, then $\mathcal{V}$ is also $U$-consistent \cite{9, 4} (and so, the the reduced dynamics is linear). Let us end this section with a proof for this statement.

 Note that some of the real coefficients $a_i$, in Eq. \eqref{eq:7}, are positive, and the others are negative. Let us denote the positive ones as $a_i^{(+)}$, and the negative ones as $a_i^{(-)}$. So, from  Eq. \eqref{eq:7}, we have
 \begin{equation}
\label{eq:14a}
\begin{aligned}
 \rho_{SE}+\sum_{i} \vert a_i^{(-)} \vert \rho_{SE}^{(i)} =\sum_{i}  a_i^{(+)}  \rho_{SE}^{(i)}+Y.
\end{aligned}
\end{equation} 
Tracing from both sides, we have $1+\sum_{i} \vert a_i^{(-)} \vert =\sum_{i} a_i^{(+)}  \equiv A$.
Dividing both sides of Eq.  \eqref{eq:14a} into $A$ results in
\begin{equation}
\label{eq:14b}
\begin{aligned}
\hat{\sigma}_{SE} =\tilde{ \sigma}_{SE}+\frac{Y}{A},
\end{aligned}
\end{equation}
where $ \hat{\sigma}_{SE}=\frac{1}{A} \left( \rho_{SE}+\sum_{i} \vert a_i^{(-)} \vert \rho_{SE}^{(i)} \right)$ and $\tilde{ \sigma}_{SE}=\frac{1}{A} \left( \sum_{i}  a_i^{(+)}  \rho_{SE}^{(i)} \right)$ are two states, on $\cH_S \otimes \cH_E$. 
Note that $ \hat{\sigma}_{SE}$ and $\tilde{ \sigma}_{SE}$ are convex combinations of the elements of the set $\mathcal{S}$. In other words, $\hat{\sigma}_{SE}, \ \tilde{ \sigma}_{SE} \in \tilde{\mathcal{S}}$, where $\tilde{\mathcal{S}}$ is the set of all convex combinations of the elements of $\mathcal{S}$, i.e., the convex hull of the set $\mathcal{S}$.

If $\tilde{\mathcal{S}}$ is $U$-consistent, then from Eqs. \eqref{eq:8a} and \eqref{eq:14b}, we conclude that Eq. \eqref{eq:14} holds, which means that $\mathcal{V}$ is, also, $U$-consistent. Obviously, when $\mathcal{V}$ is $U$-consistent, so is  $\tilde{\mathcal{S}}$,  since $\tilde{\mathcal{S}}  \subset \mathcal{V}$. Therefore, $\mathcal{V}$ is $U$-consistent if and only if $\tilde{\mathcal{S}}$ is $U$-consistent.
Consequently, for the special case that  $\mathcal{S}$ is convex, i.e.,   $\mathcal{S}=\tilde{\mathcal{S}}$, the $U$-consistency of  $\mathcal{S}$ is equivalent to  that of  $\mathcal{V}$.


\section{When the reduced dynamics is linear}  \label{sec: C}

In the previous section, we have seen that, from a convex  $U$-consistent set $\mathcal{S}$, we can construct a  $U$-consistent subspace $\mathcal{V}$, such that, for all $\rho_{SE}\in \mathcal{V}$, the reduced dynamics of the system is given by the linear Hermitian trace-preserving map  $\Phi_S$, in Eq.  \eqref{eq:16}.

In the following, we, reversely, assume that, for a set $\mathcal{S}$ and a given $U$, the reduced dynamics of the system is given by a linear (Hermitian trace-preserving) map  $\Psi_S$, and show that this assumption results that the  subspace $\mathcal{V}$, in Eq. \eqref{eq:10}, is $U$-consistent.

When the reduced dynamics of the system, for any $\rho_{S}=\mathrm{Tr}_E (\rho_{SE}), \, \rho_{SE} \in \mathcal{S}$, is given by a map $\Psi_S$, we have, from Eq. \eqref{eq:2},
\begin{equation}
\label{eq:17}
\begin{aligned}
\Psi_S(\rho_{S})=\mathrm{Tr}_{E} \circ \mathrm{Ad}_{U}(\rho_{SE}).
\end{aligned}
\end{equation} 
Assuming that $\Psi_S$ is linear, and using Eq. \eqref{eq:6}, we have
\begin{equation}
\label{eq:18}
\begin{aligned}
\Psi_S( \rho_{S})=\sum_{i=1}^m a_i \Psi_S(\rho_{S}^{(i)}),
\end{aligned}
\end{equation}
and then, using Eq. \eqref{eq:17},
\begin{equation}
\label{eq:19}
\begin{aligned}
\mathrm{Tr}_{E} \circ \mathrm{Ad}_{U}(\rho_{SE})=\sum_{i=1}^m a_i \mathrm{Tr}_{E} \circ \mathrm{Ad}_{U}(\rho_{SE}^{(i)}).
\end{aligned}
\end{equation}
Now, comparing Eqs. \eqref{eq:7} and \eqref{eq:19}, results in Eq. \eqref{eq:14}, which leads to $U$-consistency of $\mathcal{V}$, or, equivalently, $U$-consistency of  $\tilde{\mathcal{S}}$, as we have seen in the previous section.

In addition, from Eq. \eqref{eq:15}, we have $\rho_{SE}^{(i)}=\Lambda_S (\rho_{S}^{(i)})$, and so, using Eqs. \eqref{eq:6}, \eqref{eq:16},  \eqref{eq:17} and \eqref{eq:19},
\begin{equation}
\label{eq:20}
\begin{aligned}
\Psi_S(\rho_{S})=\sum_{i=1}^m a_i \mathrm{Tr}_{E} \circ \mathrm{Ad}_{U} \circ \Lambda_S (\rho_{S}^{(i)}) \\
=\mathrm{Tr}_{E} \circ \mathrm{Ad}_{U}  \circ \Lambda_S(\rho_{S})= 
\Phi_S(\rho_{S});
\end{aligned}
\end{equation}
i.e., our linear map $\Psi_S$ is the same as the linear Hermitian trace-preserving map $\Phi_S$, defined in Eq. \eqref{eq:16}.

In summary,  as our main result, we have the
following.

\begin{propo}
\label{pro:1}
Consider an arbitrary set $\mathcal{S}=\lbrace\rho_{SE}\rbrace$, of possible initial states of the system-environment. Construct the subspace $\mathcal{V}$, as in Eq.  \eqref{eq:10}. The reduced dynamics of the system, for the unitary system-environment evolution $U$, and for any initial state of the system $\rho_{S}=\mathrm{Tr}_E (\rho_{SE}), \, \rho_{SE} \in \mathcal{S}$, is given by a linear (Hermitian trace-preserving) map if and only if the  subspace $\mathcal{V}$ is $U$-consistent.
\end{propo}
In other words, the reduced dynamics is linear if and only if it can be formulated within the Dominy-Shabani-Lidar  framework, given in the previous section. Note that their framework is based on introducing a $U$-consistent $\mathcal{V}$  (and then, defining the assignment map $\Lambda_S$, as Eq. \eqref{eq:15}, and, finally, constructing the linear dynamical map $\Phi_S$, as in Eq. \eqref{eq:16}).

\begin{rem}
\label{rem:1}
During the proof of Proposition \ref{pro:1}, we have only used this fact that the system $S$ is $d_S$ dimensional, and so $0< m\leq {(d_S)}^2$. The dimension of the environment $E$ is arbitrary: $E$ can be infinite dimensional.
\end{rem}

Instead of assuming that the reduced dynamics $\Psi_S$ is linear on  $\mathcal{S}_S$ as in Eq. \eqref{eq:18}, we can assume that  $\Psi_S$ is convex-linear \cite{12a} on $\tilde{\mathcal{S}}_S=\mathrm{Tr}_E \tilde{\mathcal{S}}$. From  Eq. \eqref{eq:14b}, we have 
\begin{equation}
\label{eq:20a}
\begin{aligned}
\hat{\sigma}_{S} =\mathrm{Tr}_E(\hat{\sigma}_{SE}) =\tilde{ \sigma}_{S}=\mathrm{Tr}_E(\tilde{ \sigma}_{SE}).
\end{aligned}
\end{equation}
So, $\Psi_S(\hat{\sigma}_{S})=\Psi_S(\tilde{\sigma}_{S})$. Therefore, assuming that $\Psi_S$ is convex-linear, on $\tilde{\mathcal{S}}_S$, we have
\begin{equation}
\label{eq:20b}
\begin{aligned}
\Psi_S\left(\frac{1}{A} ( \rho_{S}+\sum_{i} \vert a_i^{(-)} \vert \rho_{S}^{(i)})\right) \qquad\qquad \\
=\Psi_S\left(\frac{1}{A} (\sum_{i} a_i^{(+)} \rho_{S}^{(i)} )\right)  \\
\Rightarrow \quad \frac{1}{A} \left(\Psi_S(\rho_{S})+\sum_{i} \vert a_i^{(-)} \vert  \Psi_S(\rho_{S}^{(i)}) \right) \qquad \\
=  \frac{1}{A}\left(\sum_{i}  a_i^{(+)}   \Psi_S(\rho_{S}^{(i)}) \right),
\end{aligned}
\end{equation}
which leads to Eq. \eqref{eq:18}.

Finally, let us summarize the results of Secs. \ref{sec: B}  and  \ref{sec: C}.
\begin{propc}{$\bf{ 1^{\prime}}$}
\label{pro:1a}
Consider an arbitrary set $\mathcal{S}=\lbrace\rho_{SE}\rbrace$, of possible initial states of the system-environment, and a given unitary time evolution of the system-environment $U$. The following statements are equivalent:
\begin{itemize}
\item[(a)] The reduced dynamics of the system, for each $\rho_{S}=\mathrm{Tr}_E (\rho_{SE}), \, \rho_{SE} \in \mathcal{S}$, is given by a linear  map.
\item[(b)] The reduced dynamics of the system, for each $\rho_{S}=\mathrm{Tr}_E (\rho_{SE}), \, \rho_{SE} \in \tilde{\mathcal{S}}$, where $\tilde{\mathcal{S}}$ is the convex hull of $\mathcal{S}$,   is given by a convex-linear  map.
\item[(c)] The set $\tilde{\mathcal{S}}$ is $U$-consistent. 
\item[(d)] The subspace $\mathcal{V}$, in Eq.  \eqref{eq:10}, is $U$-consistent.
\end{itemize}
\end{propc}
In Proposition  \ref{pro:1}, we have seen that $(a)$ implies $(d)$, and vice versa. In addition, in (the   last three paragraphs of) the previous section, it has been shown that $(c)$ implies $(d)$, and vice versa.  The reasoning given after Remark \ref{rem:1} shows that  $(b)$ implies $(a)$. Finally, $(d)$ results in Eq. \eqref{eq:16}, for each $\rho_{S}=\mathrm{Tr}_E (\rho_{SE}), \, \rho_{SE} \in \mathcal{V}$, which implies $(b)$, as a consequence.

\section{Example}  \label{sec: D}

To illustrate our results, we consider the case studied in Ref. \cite{8}. Consider a two-qubit system, one as the system $S$ and the other as the environment $E$. Assume that the Hamiltonian of the whole system-environment is \cite{8}
\begin{equation}
\label{eq:21}
\begin{aligned}
H=\frac{1}{2} \omega \, \sigma_S^{(3)} \otimes  \sigma_E^{(1)},
\end{aligned}
\end{equation}
where $\omega$ is a positive constant, and $\sigma^{(i)}$ are the Pauli operators. So, the time evolution operator, after the time interval $t$, is $U=\mathrm{exp}(-iHt)$, where $i=\sqrt{-1}$, and we set the Planck's constant $\hbar=1$.

 A general initial $\rho_{SE}$ can be expanded as
\begin{equation}
\label{eq:22}
\begin{aligned}
\rho_{SE}=\frac{1}{4} (I_{SE}+\sum_{i=1}^{3}\alpha_{i} \, \sigma_{S}^{(i)}\otimes I_{E} \qquad\qquad\qquad\quad  \\
\, +\sum_{i=1}^{3}\beta_i \, I_{S}\otimes \sigma_{E}^{(i)}
 +\sum_{i,j=1}^{3} \gamma_{ij} \, \sigma^{(i)}_{S}\otimes \sigma^{(j)}_{E}),
\end{aligned}
\end{equation}
where $I_{SE}=I_S\otimes I_E$, $I_E$ is the identity operator on $\cH_E$, and $\alpha_i , \beta_i , \gamma_{ij} \in [-1, 1]$.
So, the initial state of the system is
\begin{equation}
\label{eq:23}
\begin{aligned}
 \rho_{S}=\mathrm{Tr}_{E} (\rho_{SE})=\frac{1}{2} (I_{S}+\sum_{i=1}^{3}\alpha_{i} \, \sigma_{S}^{(i)}).
\end{aligned}
\end{equation}
The final state of the system, after the time interval $t$, using Eqs. \eqref{eq:2} and \eqref{eq:21}, is
\begin{equation}
\label{eq:24}
\begin{aligned}
\rho_{S}^\prime=\mathrm{Tr}_{E} \circ \mathrm{Ad}_{U}(\rho_{SE})=\frac{1}{2} (I_{S}+\sum_{i=1}^{3}\alpha_{i}^\prime \, \sigma_{S}^{(i)}) ,
\end{aligned}
\end{equation}
 with \cite{8}
\begin{equation}
\label{eq:25}
\begin{aligned}
\alpha_1^\prime =\alpha_1 \cos (\omega t) - \gamma_{21} \sin (\omega t), \\
\alpha_2^\prime =\alpha_2 \cos (\omega t) + \gamma_{11} \sin (\omega t), \\
\alpha_3^\prime =\alpha_3 . \qquad\qquad\qquad\qquad\quad
\end{aligned}
\end{equation}
Note that  among all the coefficients $ \gamma_{ij}$ , 
despite the symmetry between them in Eq. \eqref{eq:22},
 only 
 $ \gamma_{11}$ and  $\gamma_{21}$ are appeared, in Eq.  \eqref{eq:25}. This is due to 
the special case of the Hamiltonian, chosen in Eq. \eqref{eq:21}, which results in  Eq.  \eqref{eq:25} (after  partial tracing, over the environment $E$, as in Eq.  \eqref{eq:2}).

From Eq. \eqref{eq:25}, we can show, simply,  that when $\gamma_{11}$ and  $\gamma_{21}$ can be written as linear functions of $\alpha_i$, i.e.,  when
\begin{equation}
\label{eq:26}
\begin{aligned}
\gamma_{11}= a_{11} + \sum_{i=1}^3 b_{11}^{(i)} \alpha_i , \\
\gamma_{21}= a_{21} + \sum_{i=1}^3 b_{21}^{(i)} \alpha_i ,
\end{aligned}
\end{equation}
with real constants $a_{11}$, $a_{21}$, $ b_{11}^{(i)}$ and $ b_{21}^{(i)}$, then $\rho_{S}^\prime$, in Eq. \eqref{eq:24}, is given by a linear map from  initial  $\rho_{S}$, in Eq. \eqref{eq:23}.
Consider an initial state of the system $\rho_{S}$ as in Eq. \eqref{eq:6}, i.e., 
\begin{equation}
\label{eq:27}
\begin{aligned}
 \rho_{S}=\sum_{j=1}^m a_j \rho_{S}^{(j)}.
\end{aligned}
\end{equation}
Expand each $\rho_{S}^{(j)}\in \mathcal{S}^\prime_S$ as 
\begin{equation}
\label{eq:28}
\begin{aligned}
 \rho_{S}^{(j)}=\frac{1}{2} (I_{S}+\sum_{i=1}^{3}\alpha_{i}^{(j)} \, \sigma_{S}^{(i)}).
\end{aligned}
\end{equation}
So, using Eqs.  \eqref{eq:23}, \eqref{eq:27} and  \eqref{eq:28}, we see that
\begin{equation}
\label{eq:29}
\begin{aligned}
\alpha_i=\sum_{j=1}^{m} a_j \alpha_{i}^{(j)}. 
\end{aligned}
\end{equation}
Now, from Eqs. \eqref{eq:24}, \eqref{eq:25}, \eqref{eq:26} and \eqref{eq:29}, it is easy to show that 
\begin{equation}
\label{eq:30}
\begin{aligned}
 \rho_{S}^\prime=\sum_{j=1}^m a_j \rho_{S}^{\prime \, (j)},
\end{aligned}
\end{equation} 
where $\rho_{S}^{\prime \, (j)}=\mathrm{Tr}_{E} \circ \mathrm{Ad}_{U}(\rho_{SE}^{(j)})$ is the final state of the system, with the initial state $\rho_{S}^{(j)}$.
Therefore, defining $\Psi_S(\rho_{S}^{(i)})=\rho_{S}^{\prime \, (i)}$, we can construct a linear map 
$\Psi_S$ for which Eqs. \eqref{eq:17} and \eqref{eq:18} hold.
In summary, Eq. \eqref{eq:26} results in the existence of a linear map $\Psi_S$,  which gives the reduced dynamics of the system $S$.

Reversely, assuming that there exists a linear map $\Psi_S$, such that $\rho_{S}^\prime=\Psi_S(\rho_{S})$, results in Eq. \eqref{eq:26}.
Consider the case that $m=4$, i.e., $\mathcal{S}^\prime_S$ includes  four linear independent $\rho_{S}^{(j)}$. Let us denote the coefficient $\gamma_{11}$, for each $\rho_{SE}^{(j)}\in \mathcal{S}^\prime$, as $\gamma_{11}^{(j)}$.
In order that (the first line of) Eq. \eqref{eq:26} holds for these four $\rho_{SE}^{(j)}$, we must have
\begin{equation}
\label{eq:31}
\begin{aligned}
\left[
\begin{matrix} 
1 \quad  \alpha_1^{(1)} \quad  \alpha_2^{(1)}  \quad    \alpha_3^{(1)}  \\
1 \quad \alpha_1^{(2)} \quad  \alpha_2^{(2)} \quad     \alpha_3^{(2)}  \\
1 \quad  \alpha_1^{(3)} \quad   \alpha_2^{(3)}  \quad    \alpha_3^{(3)}  \\
1 \quad  \alpha_1^{(4)} \quad   \alpha_2^{(4)}  \quad    \alpha_3^{(4)}  
\end{matrix}
\right]
\left[
\begin{matrix} 
 a_{11}  \\   b_{11}^{(1)}   \\       b_{11}^{(2)}   \\        b_{11}^{(3)}   
 \end{matrix}
\right]=
\left[
\begin{matrix}
\gamma_{11}^{(1)}   \\     \gamma_{11}^{(2)}    \\     \gamma_{11}^{(3)}   \\     \gamma_{11}^{(4)}
\end{matrix}
\right].
\end{aligned}
\end{equation} 
Since $\rho_{S}^{(j)}$ are linearly independent, the vectors $(1,  \alpha_1^{(j)},  \alpha_2^{(j)},  \alpha_3^{(j)})$ are so. Therefore, the determinant of the first matrix, on the left hand side of Eq. 
 \eqref{eq:31}, is nonzero, and so this matrix is invertible.
Hence, we can solve Eq.  \eqref{eq:31} to find $a_{11}$ and $ b_{11}^{(i)}$ \cite{13}.
A similar line of reasoning can be given for $\gamma_{21}$. Therefore, at least for four $\rho_{SE}^{(j)}\in \mathcal{S}^\prime$, Eq. \eqref{eq:26} holds.

For any other $\rho_{SE} \in \mathcal{S}$, in general, we have
\begin{equation}
\label{eq:32}
\begin{aligned}
\gamma_{11}= a_{11} + \sum_{i=1}^3 b_{11}^{(i)} \alpha_i +\tilde{\gamma}_{11} , \\
\gamma_{21}= a_{21} + \sum_{i=1}^3 b_{21}^{(i)} \alpha_i +\tilde{\gamma}_{21}.
\end{aligned}
\end{equation}
Now, assuming that the reduced dynamics is linear, i.e., Eq. \eqref{eq:30} holds, Eqs. \eqref{eq:25} and 
 \eqref{eq:29} result that $\tilde{\gamma}_{11}=0$ and $\tilde{\gamma}_{21}=0$; i.e., for any  $\rho_{SE} \in \mathcal{S}$,  Eq.  \eqref{eq:26} holds.
In summary, the reduced dynamics of the system $S$ is linear if and only if Eq.   \eqref{eq:26} holds.

In other words, the linearity of the reduced dynamics results that the set of possible initial states of the system-environment $\mathcal{S}$ is such that  Eq.  \eqref{eq:26} holds; i.e.,  $\mathcal{S}$ includes all $\rho_{SE}$ as Eq.  \eqref{eq:22}, with arbitrary $\alpha_i$, $\beta_i$ and $\gamma_{ij}$, $(i,j)\neq (1,1), (2,1)$, but $\gamma_{11}$ and $\gamma_{21}$ are given by Eq. \eqref{eq:26}.
Note that $\mathcal{S}$ is convex. Inserting Eq. \eqref{eq:26} into Eq. \eqref{eq:25} shows that, for 
$U=\mathrm{exp}(-iHt)$,  with the Hamiltonian $H$ in Eq. \eqref{eq:21},  $\mathcal{S}$ is, also, $U$-consistent; i.e.,  for two initial $\rho_{SE}, \tau_{SE} \in \mathcal{S}$, for which we have $\rho_{S}=\mathrm{Tr}_E (\rho_{SE})=\mathrm{Tr}_E (\tau_{SE})$, the final state of the system is, also, the same.

In summary, the linearity of the reduced dynamics, i.e.,  Eq.  \eqref{eq:26}, results that the set $\mathcal{S}$ is convex and $U$-consistent, as expected from Proposition \propnumber{$ 1^{\prime}$}.


It is also worth noting that Ref. \cite{8} only considered the case that  $\gamma_{11}$ and $\gamma_{21}$ are fixed, i.e., $\gamma_{11}=a_{11}$ and $\gamma_{21}=a_{21}$, in Eq.  \eqref{eq:26}. So, Eq.  \eqref{eq:26} includes a generalization of what has been studied in Ref. \cite{8}.

\section{Nonlinearity and superluminal signaling}  \label{sec: E}

 Proposition \ref{pro:1} states that the reduced dynamics is linear if and only if  the subspace  $\mathcal{V}$, in Eq. \eqref{eq:10}, is $U$-consistent. So, if we cannot construct such a $U$-consistent
$\mathcal{V}$, from the set $\mathcal{S}$, then the reduced dynamics is not linear. It is either nonlinear or is not given by a map.

Now, an important question arises: Does the nonlinearity of the reduced dynamics result in superluminal signaling?

Gisin, in Ref. \cite{14}, considered a closed quantum system and assumed that it does not evolve linearly, as Eq.  \eqref{eq:1}. He proposed a \textit{gedanken} nonlinear evolution model. For that model, he
 showed that the nonlinear evolution leads to superluminal signaling; i.e., after the evolution, one can perform measurements, on that closed quantum system, such that the results of those measurements lead to  superluminal communications. 

However,  assuming that the linear dynamics, for a closed quantum system, as Eq. \eqref{eq:1}, does not lead to superluminal signaling means one can perform no measurement, on such a system, which results in superluminal communications. 
One kind of measurements, which one can perform on a system, are those that can be done on a subsystem of the whole  system, i.e., those which are determined knowing the reduced density operator of this subsystem.
 Obviously, for this restricted class of measurements, no  superluminal signaling occurs.

We can use the above argument for the whole system-environment, which is a closed quantum system and evolves linearly, as Eq. \eqref{eq:1}: performing  measurements on $S$ cannot lead to superluminal communications, regardless of  whether the reduced dynamics of $S$ is linear, or not.

Let us  emphasize again that the (non)linearity of the reduced dynamics is only a consequence of $U$-(in)consistency of  $\mathcal{V}$, while the dynamics of the whole system-environment is linear. It differs, fundamentally, from the Gisin's example, in which the dynamics (of a closed system) is, itself,   nonlinear.

Now, we can follow two different points of view: first, we may consider the quantum theory as a theory of preparation, evolution and  measurement \cite{15}. So, since the preparation is a part of the theory, 
 $U$-(in)consistency of initial  $\mathcal{V}$ is a part of the theory, which determines the (non)linearity of the (reduced) dynamics.

Second, we may consider the evolution (and the measurement) physical, i.e., as parts of the physics (theory), but not the preparation.
From this point of  view, the (non)linearity of the reduced dynamics, as a consequence of $U$-(in)consistency of initial $\mathcal{V}$, does not seem rather physical. This may be the reason that the authors of Ref. \cite{16} proposed a different approach to the dynamics of open quantum
systems, which they argued  is more causal. However, we think that this issue needs more consideration.

Let us end this section, with
the following point. We have seen that the nonlinear reduced dynamics cannot lead to superluminal communications. But, even when the reduced dynamics is linear, some other  unexpected results may occur. 
For example, it is known that the \textit{trace distance} \cite{18a}, between two states, does not increase, under completely positive maps \cite{1}. But, when the (reduced) dynamics is not completely positive, this  contractivity property may be violated. 

 In Ref. \cite{18}, a two-qubit case, one as the system $S$ and the other as the environment $E$, is considered. The system-environment  evolution is given by the \textit{swap} operator  $U_{sw}$, where $U_{sw}\vert\psi\rangle\vert\phi\rangle=\vert\phi\rangle\vert\psi\rangle$.
 By choosing an appropriate subspace  $\mathcal{V}$  (let us denote it as  $\mathcal{V}_1$), it has been shown that the reduced dynamics is given by a linear Hermitian trace-preserving map, which is non-positive and is such that the trace distance increases, after the evolution \cite{18}. (There, in Ref. \cite{18}, that non-positive map is called the  \textit{repolarizer} map.)

Let us instead  choose  
\begin{equation}
 \label{eq:five-a}
\begin{aligned}
\mathcal{S}=\mathcal{S}_2=\lbrace\rho_{SE}=\rho_{S}\otimes\tilde{\omega}_{E}\rbrace \qquad \\
\Rightarrow \quad \mathcal{V}=\mathcal{V}_2= \mathrm{Span}_{\mathbb{C}} \  \mathcal{S}_2 \neq \mathcal{V}_1, 
\end{aligned}
\end{equation} 
where $\rho_{S}$ are arbitrary states of the system, but $\tilde{\omega}_{E}$ is a fixed state of the environment. Now, for this $\mathcal{V}_2$, and any arbitrary  system-environment evolution $U$, in our case  $U=U_{sw}$,  the reduced dynamics is completely positive \cite{1}, and so, the trace distance is contractive. Therefore, the contractivity of the trace distance, for each two initial states of the system $\rho_S, \sigma_S \in \mathrm{Tr}_{E} (\mathcal{V}_2 \cap \mathcal{D}_{SE})= \mathrm{Tr}_{E} \mathcal{S}_2=\mathcal{D}_S$ (where $\mathcal{D}_S$ and $\mathcal{D}_{SE}$  are the sets of all states on $\cH_S$ and $\cH_S\otimes\cH_E$, respectively),  
is only a consequence of choosing initial $\mathcal{V}_2$ as Eq. \eqref{eq:five-a}, and is not related to the system-environment evolution  $U=U_{sw}$. So, following the second point of view, this contractivity property cannot be considered rather physical, though, it is valid for each two initial states $\rho_S, \sigma_S \in \mathcal{D}_S$.


\qquad \\

\section{Summary} \label{sec: F}

In Ref. \cite{4}, it has been shown that a $U$-consistent subspace $\mathcal{V}$ results in linear reduced dynamics . In this paper, we showed that the reverse is, also, true: linear reduced dynamics results in the $U$-consistency of the subspace $\mathcal{V}$, in Eq. \eqref{eq:10}.

To illustrate this result, in Sec. \ref{sec: D}, we considered a two-qubit case,  studied in \cite{8}, one as the system $S$ and the other as the environment $E$,  and showed that how the linearity of the reduced dynamics of $S$, i.e., Eq. \eqref{eq:26}, leads to the $U$-consistency of $\mathcal{V}$.
Studying other examples, especially with higher dimensional $S$ or $E$,  can  help  illustrate Proposition \ref{pro:1} further.

Finally, in Sec. \ref{sec: E}, we have seen that the nonlinearity of the reduced dynamics cannot lead to the 
superluminal signaling. This is, however, an expected result; since we do not expect that the properties of the set $\mathcal{S}$ (the subspace $\mathcal{V}$) affect the (im)possibility of superluminal signaling.


\subsection*{Acknowledgments}

I would like to thank the two anonymous referees for their helpful comments.


\begin{thebibliography}{1}
\bibitem{12a} When $\Psi_S$ is  convex-linear, on $\tilde{\mathcal{S}}_S$, then we have  $\Psi_S \left( p \rho_{S}+ (1-p) \tau_{S}\right)=p\Psi_S(\rho_{S})+(1-p)\Psi_S(\tau_{S})$, where $\rho_{S}, \ \tau_{S} \in \tilde{\mathcal{S}}_S$ and  $0\leq p\leq 1$.
\bibitem{13} When $m<4$, we can proceed with the same line of reasoning for a  set of four linearly independent states $\rho_{S}^{(j)}$, which includes $\mathcal{S}_S^\prime$ as a subset.
\bibitem{18a} The trace distance, between two states $\rho$ and $\sigma$, is defined as $\frac{1}{2}\mathrm{Tr}(\vert\rho - \sigma\vert)$, where $\vert B \vert$, for an operator $B$, is $\sqrt{B^{\dagger}B}$ \cite{1}.
\end{thebibliography}
\end{document}